\documentclass[prd,aps,nofootinbib,eqsecnum]{revtex4}
\usepackage{graphicx,color,amsmath,amsxtra}
\usepackage{amssymb}
\usepackage{enumerate}
\usepackage{hhline}
\usepackage{array}
\usepackage{tabularx}

\newcommand{\bear}{\begin{array}}  \newcommand{\eear}{\end{array}}
\newcommand{\bea}{\begin{eqnarray}}  \newcommand{\eea}{\end{eqnarray}}
\newcommand{\beq}{\begin{equation}}  \newcommand{\eeq}{\end{equation}}
\newcommand{\bef}{\begin{figure}}  \newcommand{\eef}{\end{figure}}
\newcommand{\bec}{\begin{center}}  \newcommand{\eec}{\end{center}}

\newcommand{\Eqn}[1]{&\hspace{-0.2em}#1\hspace{-0.2em}&}

\def\Vec#1{\mbox{\boldmath $#1$}}

\def\be{\begin{equation}}
\def\ee{\end{equation}}
\def\bea{\begin{eqnarray}}
\def\eea{\end{eqnarray}}
\def\beq{\begin{eqnarray}}
\def\eeq{\end{eqnarray}}

\begin{document}

\title{Finite-time future singularities in modified gravity
}

\author{
Kazuharu Bamba\footnote{E-mail address: bamba``at"phys.nthu.edu.tw}}
\affiliation{
Department of Physics, National Tsing Hua University, Hsinchu, Taiwan 300
}


\begin{abstract}
We review finite-time future singularities in modified gravity. 
We reconstruct an explicit model of modified gravity realizing a crossing of 
the phantom divide and show that the Big Rip singularity appears in the 
modified gravitational theory. It is also demonstrated that 
the (finite-time) Big Rip singularity in the modified gravity is transformed 
into the infinite-time singularity in the corresponding scalar field theory 
obtained through the conformal transformation. 
Furthermore, we study several models of modified gravity which 
produce accelerating cosmologies ending at the finite-time future 
singularities of all four known types. 
\end{abstract}


\maketitle

\section{Introduction}

Recent observations confirmed that the current expansion of the universe 
is accelerating. 
There are two approaches to explain the current accelerated expansion of 
the universe. One is to introduce some unknown matter, which is called 
``dark energy'' in the framework of general relativity. 
The other is to modify the gravitational theory, e.g., to study the action 
described by an arbitrary function of the scalar
curvature $R$. This is called ``$F(R)$ gravity'', where 
$F(R)$ is an arbitrary function of the scalar curvature $R$ 
(for reviews, see~\cite{Nojiri:2006ri, review-2}). 

According to the recent various observational data, 
there exists the possibility that the effective equation of state (EoS)
parameter, which is the ratio of the effective pressure of the universe to
the effective energy density of it, 
evolves from larger than $-1$ (non-phantom phase) to less than $-1$
(phantom one, in which superacceleration is realized), namely, 
crosses $-1$ (the phantom divide) currently or in near future. 
Various attempts to realize the crossing of the phantom divide have been 
made in the framework of general relativity. 
Recently, a crossing of the phantom divide in modified gravity has also been 
investigated~\cite{Nojiri:2006ri, Abdalla:2004sw, Amendola:2007nt, Bamba:2008hq}. 
Moreover, it is known that modified gravity may lead to the effective 
phantom/quintessence phase~\cite{Nojiri:2006ri}, while the phantom/quintessence-dominated universe may end up with finite-time future singularities, 
which can be categorized into four types~\cite{Nojiri:2005sx}. 

In the present article, 
we review finite-time future singularities in modified gravity. 
Following the considerations in Ref.~\cite{Bamba:2008hq}, 
we reconstruct an explicit model of modified gravity realizing a crossing of 
the phantom divide by using the reconstruction method proposed in 
Refs.~\cite{Nojiri:2006gh, reconstruction method-2}. 
We show that the Big Rip singularity appears in this modified gravitational 
theory, 
whereas that the (finite-time) Big Rip singularity in the modified gravity is 
transformed into the infinite-time singularity in the corresponding scalar 
field theory. 
Next, following the investigations in Ref.~\cite{Bamba:2008ut}, 
we explore several examples of $F(R)$ gravity which 
predict the accelerating cosmological solutions ending at the finite-time 
future singularities. It is demonstrated that not only the 
Big Rip but other three types of the finite-time future singularities 
may appear. 

This article is organized as follows. 
In Sec.~II we explain the reconstruction method of modified 
gravity~\cite{Nojiri:2006gh, reconstruction method-2}. Using this method, 
we reconstruct an explicit model of modified gravity in which 
a crossing of the phantom divide can be realized. 
We also consider the corresponding scalar field theory. 
In Sec.~III we present several models of $F(R)$ gravity which predict 
accelerating cosmologies ending at the finite-time future singularities by 
using the reconstruction method. 
Finally, summary is given in Sec.~IV. 
We use units in which $k_\mathrm{B} = c = \hbar = 1$ and denote the
gravitational constant $8 \pi G$ by ${\kappa}^2$, so that
${\kappa}^2 \equiv 8\pi/{M_{\mathrm{Pl}}}^2$, where
$M_{\mathrm{Pl}} = G^{-1/2} = 1.2 \times 10^{19}$GeV is the Planck mass. 

\section{Modified gravitational theory realizing 
a crossing of the phantom divide}

We investigate modified gravity in which a crossing of 
the phantom divide can be realized by using the reconstruction method. 

\subsection{Reconstruction of modified gravity}

First, we review the reconstruction method of modified 
gravity proposed in Refs.~\cite{Nojiri:2006gh, reconstruction method-2} 
(for the related study of reconstruction in $F(R)$ gravity, 
see~\cite{Cortes:2008fy}). 

The action of $F(R)$ gravity with general matter is given by
\begin{eqnarray}
S = \int d^4 x \sqrt{-g} \left[ \frac{F(R)}{2\kappa^2} +
{\mathcal{L}}_{\mathrm{matter}} \right]\,,
\label{eq:2.1}
\end{eqnarray}
where $g$ is the determinant of the metric tensor $g_{\mu\nu}$ and
${\mathcal{L}}_{\mathrm{matter}}$ is the matter Lagrangian.

The action (\ref{eq:2.1}) can be rewritten to the following form
by using proper functions $P(\phi)$ and $Q(\phi)$ of a scalar field $\phi$:
\begin{eqnarray}
S=\int d^4 x \sqrt{-g} \left\{ \frac{1}{2\kappa^2} \left[ P(\phi) R + Q(\phi)
\right] + {\mathcal{L}}_{\mathrm{matter}} \right\}\,.
\label{eq:2.2}
\end{eqnarray}
The scalar field $\phi$ may be regarded as an auxiliary scalar field because
$\phi$ has no kinetic term. Taking the variation of the action (\ref{eq:2.2})
with respect to $\phi$, we obtain
\begin{eqnarray}
0=\frac{d P(\phi)}{d \phi} R + \frac{d Q(\phi)}{d \phi}\,,
\label{eq:2.3}
\end{eqnarray}
which may be solved with respect to $\phi$ as $\phi=\phi(R)$.
Substituting $\phi=\phi(R)$ into the action (\ref{eq:2.2}),
we find that the expression of $F(R)$ in the action of $F(R)$ gravity in
Eq.~(\ref{eq:2.1}) is given by
\begin{eqnarray}
F(R) = P(\phi(R)) R + Q(\phi(R))\,.
\label{eq:2.4}
\end{eqnarray}

Taking the variation of the action (\ref{eq:2.2}) with respect to the metric
$g_{\mu\nu}$, we find that the field equation of modified gravity
is given by
\begin{eqnarray}
\frac{1}{2}g_{\mu \nu} \left[ P(\phi) R + Q(\phi) \right]
-R_{\mu \nu} P(\phi) -g_{\mu \nu} \Box P(\phi) +
{\nabla}_{\mu} {\nabla}_{\nu}P(\phi) + \kappa^2
T^{(\mathrm{matter})}_{\mu \nu} = 0\,,
\label{eq:2.5}
\end{eqnarray}
where ${\nabla}_{\mu}$ is the covariant derivative operator associated with
$g_{\mu \nu}$, $\Box \equiv g^{\mu \nu} {\nabla}_{\mu} {\nabla}_{\nu}$
is the covariant d'Alembertian for a scalar field, and
$T^{(\mathrm{matter})}_{\mu \nu}$ is the contribution to
the matter energy-momentum tensor.

We assume the flat 
Friedmann-Robertson-Walker (FRW) space-time with the metric 
$
{ds}^2 = -{dt}^2 + a^2(t)d{\Vec{x}}^2
$, 
where $a(t)$ is the scale factor. 
In this background, the $(\mu,\nu)=(0,0)$ component and
the trace part of the $(\mu,\nu)=(i,j)$ component of Eq.~(\ref{eq:2.5}),
where $i$ and $j$ run from $1$ to $3$, read
\begin{eqnarray}
&&
-6H^2P(\phi(t)) -Q(\phi(t)) -6H \frac{dP(\phi(t))}{dt} + 2\kappa^2\rho = 0\,,
\label{eq:2.7} \\
&&
\hspace{-40.5mm}
\mathrm{and} 
\hspace{25mm}
2\frac{d^2P(\phi(t))}{dt^2}+4H\frac{dP(\phi(t))}{dt}+
\left(4\dot{H}+6H^2 \right)P(\phi(t)) +Q(\phi(t)) + 2\kappa^2 p = 0\,,
\label{eq:2.8}
\end{eqnarray}
respectively,
where $H=\dot{a}/a$ is the Hubble parameter.
Here, $\rho$ and $p$ are the sum of the energy density and
pressure of matters with a constant
EoS parameter $w_i$, respectively, where $i$ denotes some component of
the matters.

Eliminating $Q(\phi)$ from Eqs.~(\ref{eq:2.7}) and (\ref{eq:2.8}), we obtain
\begin{eqnarray}
\frac{d^2P(\phi(t))}{dt^2} -H\frac{dP(\phi(t))}{dt} +2\dot{H}P(\phi(t)) +
\kappa^2 \left( \rho + p \right) = 0\,.
\label{eq:2.9}
\end{eqnarray}
We note that the scalar field $\phi$ may be taken as $\phi = t$ because
$\phi$ can be redefined properly.

We now consider that $a(t)$ is described as 
$
a(t) = \bar{a} \exp \left( \tilde{g}(t) \right)
$, 
where
$\bar{a}$ is a constant and $\tilde{g}(t)$ is a proper function.
In this case, Eq.~(\ref{eq:2.9}) is reduced to
\begin{eqnarray}
&&
\frac{d^2P(\phi)}{d\phi^2} -\frac{d \tilde{g}(\phi)}{d\phi}
\frac{dP(\phi)}{d\phi} +2 \frac{d^2 \tilde{g}(\phi)}{d \phi^2}
P(\phi) 
+\kappa^2 \sum_i \left( 1+w_i \right) \bar{\rho}_i
\bar{a}^{-3\left( 1+w_i \right)} \exp
\left[ -3\left( 1+w_i \right) \tilde{g}(\phi) \right] = 0\,,
\label{eq:2.11}
\end{eqnarray}
where $\bar{\rho}_i$ is a constant and we have used
$H= d \tilde{g}(\phi)/\left(d \phi \right)$.
Moreover, it follows from Eq.~(\ref{eq:2.7}) that $Q(\phi)$ is given by 
\begin{eqnarray}
Q(\phi) \Eqn{=} -6 \left[ \frac{d \tilde{g}(\phi)}{d\phi} \right]^2 P(\phi)
-6\frac{d \tilde{g}(\phi)}{d\phi} \frac{dP(\phi)}{d\phi} 
+2\kappa^2 \sum_i \bar{\rho}_i \bar{a}^{-3\left( 1+w_i \right)}
\exp
\left[ -3\left( 1+w_i \right) \tilde{g}(\phi) \right]\,.
\label{eq:2.12}
\end{eqnarray}
Hence, if we obtain the solution of Eq.~(\ref{eq:2.11}) with respect to
$P(\phi)$, then we can find $Q(\phi)$. Consequently, using Eq.~(\ref{eq:2.4}),
we can reconstruct $F(R)$ gravity for any cosmology expressed by 
$
a(t) = \bar{a} \exp \left( \tilde{g}(t) \right)
$. 
In Refs.~\cite{Nojiri:2006gh, U-S, Nojiri:2007as}, 
specific models unifying the sequence: the early-time acceleration, 
radiation/matter-dominated stage and dark energy epoch have been constructed.


Next, using the above method, we reconstruct an explicit model in which 
a crossing of the phantom divide can be realized. 
We start with Eq.~(\ref{eq:2.11}) without matter: 
\be
\label{PDF1}
\frac{d^2 P(\phi)}{d\phi^2} - \frac{d \tilde{g}(\phi)}{d\phi}
\frac{d P(\phi)}{d\phi}
+ 2 \frac{d^2 \tilde{g}(\phi)}{d\phi^2} P(\phi) = 0\ .
\ee
By redefining $P(\phi)$ as 
$
P(\phi) = e^{\tilde{g}(\phi)/2} \tilde{p}(\phi) 
$, 
Eq.~(\ref{PDF1}) is rewritten to
\be
\label{PDF3}
\frac{1}{\tilde{p}(\phi)} \frac{d^2 \tilde{p}(\phi)}{d \phi^2}
= 25 e^{\tilde{g}(\phi)/10} \frac{d^2 \left(e^{-\tilde{g}(\phi)/10}\right)}
{d\phi^2}\ .
\ee
We explore the following model: 
$
\tilde{g}(\phi) = - 10 \ln \left[ \left(\phi/t_0\right)^{-\gamma} 
 - C \left(\phi/t_0\right)^{\gamma+1} \right]
$, 
where $C$ and $\gamma$ are positive constants,
and $t_0$ is the present time.
In this case, Eq.~(\ref{PDF3}) is reduced to 
$
\left(1/\tilde{p}(\phi)\right) 
\left[ d^2 \tilde{p}(\phi)/\left(d \phi^2\right)\right]
= 25 \gamma (\gamma + 1)/\phi^2
$, 
which can be solved as 
$
\tilde{p}(\phi) = \tilde{p}_+ \phi^{\beta_+} + \tilde{p}_- \phi^{\beta_-}
$. 
Here, $\tilde{p}_\pm$ are arbitrary constants and $\beta_\pm$ are given by 
$
\beta_\pm = \left[1 \pm \sqrt{1 + 100 \gamma (\gamma + 1)} \right]/2
$. 
 From the above expression of $\tilde{g}(\phi)$, we find that 
$\tilde{g}(\phi)$ diverges at finite $\phi$ when 
$
\phi = t_s \equiv t_0 C^{-1/(2\gamma + 1)}
$, 
which tells that there could be the Big Rip singularity at $t=t_s$. 
We only need to consider the period $0<t<t_s$ because 
$\tilde{g}(\phi)$ should be real number. 
In this case, the Hubble rate $H(t)$ is given by 
\be
\label{PDF9}
H(t)= \frac{d \tilde{g}(\phi)}{d \phi}
= \left(\frac{10}{t_0}\right) \left[ \gamma \left(\frac{\phi}{t_0} 
\right)^{-\gamma-1 } 
+ (\gamma+1) C \left(\frac{\phi}{t_0}\right)^{\gamma} \right] 
\left[ \left(\frac{\phi}{t_0}\right)^{-\gamma}
 - C \left(\frac{\phi}{t_0}\right)^{\gamma+1} \right]^{-1}\ ,
\ee
where it is taken $\phi=t$.

In the flat FRW background, even for modified gravity described by 
the action (\ref{eq:2.1}), the effective energy-density
and pressure of the universe are given by
$\rho_\mathrm{eff} = 3H^2/\kappa^2$ and
$p_\mathrm{eff} = -\left(2\dot{H} + 3H^2 \right)/\kappa^2$, respectively.
The effective EoS parameter
$w_\mathrm{eff} = p_\mathrm{eff}/\rho_\mathrm{eff}$
is defined as 
$
w_\mathrm{eff} \equiv -1 -2\dot{H}/\left(3H^2\right)
$~\cite{Nojiri:2006ri}. 
For the case of $H(t)$ in Eq.~(\ref{PDF9}), $w_\mathrm{eff}$ is expressed as 
$
w_\mathrm{eff} = -1 + U(t)
$, 
where
\begin{eqnarray}
U(t) \equiv -\frac{2\dot{H}}{3H^2} =
- \frac{1}{15} 
\left[ -\gamma + 4\gamma \left( \gamma+1 \right)
\left( \frac{t}{t_s} \right)^{2\gamma+1} +
\left( \gamma+1 \right) \left( \frac{t}{t_s}
\right)^{2\left( 2\gamma+1 \right)} \right]
\left[ \gamma + \left( \gamma+1 \right)
\left( \frac{t}{t_s} \right)^{2\gamma+1} \right]^{-2}\,.
\label{eq:I0-1-2}
\end{eqnarray}
For the case of Eq.~(\ref{PDF9}), 
the scalar curvature $R=6\left( \dot{H} + 2H^2 \right)$ is expressed as 
\begin{eqnarray}
\hspace{-5mm}
R =
60
\Biggl[
\gamma \left( 20\gamma -1 \right) + 44\gamma \left( \gamma+1 \right)
\left( \frac{t}{t_s} \right)^{2\gamma+1}
+ \left( \gamma+1 \right) \left( 20\gamma+21 \right)
\left( \frac{t}{t_s} \right)^{2\left( 2\gamma+1 \right)}
\Biggr]
t^{-2} \left[ 1- \left( \frac{t}{t_s} \right)^{2\gamma+1} \right]^{-2}
\,.
\label{eq:I0-2}
\end{eqnarray}
In deriving Eqs.~(\ref{eq:I0-1-2}) and (\ref{eq:I0-2}), we have used 
$
t_s = t_0 C^{-1/(2\gamma + 1)}
$. 

When $t\to 0$, i.e., $t \ll t_s$, $H(t)$ behaves as 
$
H(t) \sim 10\gamma/t
$. 
In this limit, it follows from 
$
w_\mathrm{eff} = -1 -2\dot{H}/\left(3H^2\right)
$ 
that 
the effective EoS parameter is given by 
$
w_\mathrm{eff} = -1 + 1/\left(15\gamma\right)
$. 
This behavior is 
identical with that in the Einstein gravity with matter
whose EoS parameter is greater than $-1$. 

On the other hand, when $t\to t_s$, we find 
$
H(t) \sim 10/\left(t_s - t\right)
$. 
In this case, the scale factor is given by
$a(t) \sim \bar{a} \left( t_s - t \right)^{-10}$.
When $t\to t_s$, therefore, $a \to \infty$, namely, the Big Rip singularity
appears.
In this limit, the effective EoS parameter is given by 
$
w_\mathrm{eff} = - 1 - 1/15 = -16/15
$. 
This behavior is identical with the case in which there is a phantom matter
with its EoS parameter being smaller than $-1$. 
Thus, we have obtained an explicit model showing a crossing of
the phantom divide.

It follows from 
$
w_\mathrm{eff} = -1 -2\dot{H}/\left(3H^2\right)
$ 
that the effective EoS parameter
$w_\mathrm{eff}$ becomes $-1$ when $\dot{H}=0$.
Solving $w_\mathrm{eff} = -1$ with respect to 
$t$ by using 
$
w_\mathrm{eff} = -1 + U(t)
$, 
namely, $U(t)=0$, 
we find that the effective EoS parameter crosses the phantom divide at
$t=t_\mathrm{c}$ given by
$ 
t_\mathrm{c} = t_s \left[ -2\gamma +
\sqrt{4\gamma^2 + \gamma/\left(\gamma+1\right)} \,
\right]^{1/\left( 2\gamma + 1 \right)}
$. 
 From Eq.~(\ref{eq:I0-1-2}), we see that when $t<t_\mathrm{c}$, $U(t)>0$
because $\gamma >0$.
Moreover, the time derivative of $U(t)$ is given by
\begin{eqnarray}
\frac{d U(t)}{dt} = 
-\frac{1}{15} 
\left[2\gamma \left( \gamma+1 \right) \left( 2\gamma+1 \right)^2\right] 
\left[ \gamma + \left( \gamma+1 \right)
\left( \frac{t}{t_s} \right)^{2\gamma+1} \right]^{-3} 
\left( \frac{1}{t_s} \right)
\left( \frac{t}{t_s} \right)^{2\gamma}
\left[ 1 - \left( \frac{t}{t_s} \right)^{2\gamma+1}
\right]\,.
\label{eq:I1-2}
\end{eqnarray}
Eq.~(\ref{eq:I1-2}) tells that the relation $d U(t)/\left(dt\right) <0$ is
always satisfied because we only consider the period $0<t<t_s$ as mentioned
above.
This means that $U(t)$ decreases monotonously. Thus, the value of $U(t)$
evolves from positive to negative. From 
$
w_\mathrm{eff} = -1 + U(t)
$, 
we see that 
the value of $w_\mathrm{eff}$ crosses $-1$.
Once the universe enters the phantom phase, it stays in this phase,
namely, the value of $w_\mathrm{eff}$ remains less than $-1$, and
finally the Big Rip singularity appears because $U(t)$ decreases
monotonically. 

As a consequence, $P(t)$ is given by 
$
P(t) = \left\{ \left( t/t_0 \right)^\gamma/ 
\left[1-\left( t/t_s \right)^{2\gamma+1}\right] \right\}^5 
\sum_{j=\pm} \tilde{p}_j t^{\beta_j}
$. 
Using Eqs.~(\ref{eq:2.12}), we obtain 
$
Q(t) = -6H
\left\{ \left( t/t_0 \right)^\gamma/ 
\left[1-\left( t/t_s \right)^{2\gamma+1}\right] \right\}^5
\sum_{j=\pm} \left( 3H/2 + \beta_j/t \right) 
\tilde{p}_j t^{\beta_j} 
$. 

When $t\to 0$, from   
$
H(t) \sim 10\gamma/t
$, 
we find
$
t \sim \sqrt{60\gamma \left( 20\gamma -1 \right)/R}
$. 
In this limit, it follows from Eqs.~(\ref{eq:2.4}) that 
the form of $F(R)$ is given by
\begin{eqnarray}
\hspace{-10mm}
F(R) \Eqn{\sim}
\left\{
\frac{\left[\frac{1}{t_0} \sqrt{60\gamma \left( 20\gamma -1 \right)} R^{-1/2}
\right]^\gamma}{1 -
\left[\frac{1}{t_s} \sqrt{60\gamma \left( 20\gamma -1 \right)} R^{-1/2}
\right]^{2\gamma+1}} \right\}^5 R 
\sum_{j=\pm}
\biggl\{ \left( \frac{5\gamma -1 -\beta_j}{20\gamma -1} \right) \tilde{p}_j
\left[60\gamma \left( 20\gamma -1 \right) \right]^{\beta_j /2}
R^{-\beta_j /2}
\biggr\}\,.
\label{eq:I5}
\end{eqnarray}

On the other hand, when $t\to t_s$, from 
$
H(t) \sim 10/\left(t_s - t\right)
$, 
we obtain 
$
t \sim t_s - 3\sqrt{140/R}
$. 
In this limit, it follows from Eqs.~(\ref{eq:2.4}) that 
the form of $F(R)$ is given by
\begin{eqnarray}
F(R) \Eqn{\sim}
\left[ 
\frac{\left( J/t_0 \right)^\gamma}{1 - \left(J/t_s\right)^{2\gamma+1}} 
\right]^5 R 
\sum_{j=\pm}
\tilde{p}_j J^{\beta_j} 
\left\{
1- \sqrt{\frac{20}{7}}
\left[
\sqrt{\frac{15}{84}} t_s
+ \left( \beta_j - 15 \right) R^{-1/2}
\right]\frac{1}{J}
\right\}\,, 
\label{eq:I7}
\end{eqnarray}
where $J \equiv t_s - 3\sqrt{140/R}$. 
The above modified gravity may be considered as some approximated form of 
more realistic, viable theory. 
 From Eq.~(\ref{eq:I0-2}), we see that in the above limit the scalar curvature
diverges, and that the expression of $F(R)$ in (\ref{eq:I7}) can be
approximately written as
\begin{eqnarray}
F(R) \approx
\frac{2}{7}
\left[
\frac{1}{3\sqrt{140} \left( 2\gamma +1 \right)}
\left( \frac{t_s}{t_0} \right)^\gamma \right]^5
\left(
\sum_{j=\pm}
\tilde{p}_j t_s^{\beta_j} \right)
t_s^5 R^{7/2}\,.
\label{eq:I8}
\end{eqnarray}

\subsection{Corresponding scalar field theory}

In this subsection, motivated by the discussion in 
Ref.~\cite{Briscese:2006xu}, 
we consider the corresponding scalar field theory to modified gravity
realizing a crossing of the phantom divide, which is obtained 
by making the conformal transformation of the modified gravitational
theory. (In Ref.~\cite{Capozziello:2005mj}, the relations between scalar field 
theories and $F(R)$ gravity have been studied.)
By introducing two scalar fields $\zeta$ and $\xi$, we can rewrite
the action~(\ref{eq:2.1}) to the following form~\cite{Nojiri:2006ri}:
\begin{eqnarray}
S \Eqn{=}
\int d^{4}x \sqrt{-g}
\left\{
\frac{1}{2\kappa^2} \left[ \xi \left( R-\zeta \right) + F(\zeta) \right] +
{\mathcal{L}}_{\mathrm{matter}}
\right\}\,.
\label{eq:3.1}
\end{eqnarray}
The form in Eq.~(\ref{eq:3.1}) is reduced to the original expression 
in Eq.~(\ref{eq:2.1}) by using the equation $\zeta=R$, which is derived
by taking variation of the action~(\ref{eq:3.1}) with respect to
one auxiliary field $\xi$.
Taking the variation of the form in Eq.~(\ref{eq:3.1}) with respect
to the other auxiliary field $\zeta$, we obtain 
$
\xi = F^{\prime}(\zeta)
$, 
where the prime denotes differentiation with respect to $\zeta$.
Substituting this equation into Eq.~(\ref{eq:3.1}) and eliminating 
$\xi$ from Eq.~(\ref{eq:3.1}), we find 
\begin{eqnarray}
S \Eqn{=}
\int d^{4}x \sqrt{-g}
\left[
\frac{1}{2\kappa^2} \left(
F^{\prime}(\zeta) R + F(\zeta) - F^{\prime}(\zeta) \zeta \right) +
{\mathcal{L}}_{\mathrm{matter}}
\right]\,.
\label{eq:3.3}
\end{eqnarray}
This is the action in the Jordan-frame, in which there exists a non-minimal
coupling between $F^{\prime}(\zeta)$ and the scalar curvature $R$.
We make the following conformal transformation of the action (\ref{eq:3.3}): 
$
g_{\mu \nu} \rightarrow  
\hat{g}_{\mu \nu} = e^{\sigma} g_{\mu \nu}
$, 
where 
$
e^{\sigma} = F^{\prime}(\zeta)
$. 
Here, $\sigma$ is a scalar field and a hat denotes quantities
in the Einstein frame, in which the non-minimal coupling between
$F^{\prime}(\zeta)$ and the scalar curvature $R$ in the first term on the
right-hand side of Eq.~(\ref{eq:3.3}) disappears.
Consequently, the action in the Einstein frame is given
by~\cite{F-M, Nojiri:2003ft}
\begin{eqnarray}
S_{\mathrm{E}} =
\int d^{4}x \sqrt{-\hat{g}}
\left[
\frac{1}{2\kappa^2} \left( \hat{R} - \frac{3}{2} \hat{g}^{\mu\nu}
{\partial}_{\mu} \sigma {\partial}_{\nu} \sigma
-V(\sigma) \right)
+ e^{-2\sigma} {\mathcal{L}}_{\mathrm{matter}}
\right]\,, 
\label{eq:3.6} 
\end{eqnarray}
where 
$
V(\sigma) = e^{-\sigma} \zeta(\sigma) - e^{-2\sigma}
F\left( \zeta (\sigma) \right)
= \zeta/F^{\prime}(\zeta) - F(\zeta)/\left( F^{\prime}(\zeta) \right)^2 
$
and $\hat{g}$ is the determinant of $\hat{g}^{\mu\nu}$. 
In deriving Eqs.~(\ref{eq:3.6}), we have used 
$
e^{\sigma} = F^{\prime}(\zeta)
$. 
In addition, $\zeta(\sigma)$ is obtained by solving 
$
e^{\sigma} = F^{\prime}(\zeta)
$ 
with respect to $\zeta$ as $\zeta=\zeta(\varphi)$.
Defining $\varphi$ as $\varphi \equiv \sqrt{3/2} \sigma/\kappa$,
the action (\ref{eq:3.6}) is reduced to the following form of the canonical
scalar field theory:
\begin{eqnarray}
S_{\mathrm{ST}} =
\int d^{4}x \sqrt{-\hat{g}}
\left[
\frac{\hat{R}}{2\kappa^2} - \frac{1}{2} \hat{g}^{\mu\nu}
{\partial}_{\mu} \varphi {\partial}_{\nu} \varphi
-V(\varphi)
+ e^{-2\sqrt{2/3} \kappa \varphi} {\mathcal{L}}_{\mathrm{matter}}
\right]\,.
\label{eq:3.8}
\end{eqnarray} 

Taking the variation of the action (\ref{eq:2.1}) with respect to the metric
$g_{\mu\nu}$, we find that the field equation of modified gravity
is given by 
$
F^{\prime}(R) R_{\mu \nu}
- \left(1/2\right) g_{\mu \nu} F(R) + g_{\mu \nu}
\Box F^{\prime}(R) - {\nabla}_{\mu} {\nabla}_{\nu} F^{\prime}(R)
= \kappa^2 T^{(\mathrm{matter})}_{\mu \nu}
$. 
When there is no matter, using the $(\mu,\nu)=(0,0)$ component and 
the trace part of the $(\mu,\nu)=(i,j)$ component of the above gravitational 
field equation, in the flat FRW background, we obtain 
\begin{eqnarray}
\hspace{-5mm}
2\dot{H}F^{\prime}(R)
+6\left( -4H^2 \dot{H} +4\dot{H}^2 + 3H\ddot{H} + \dddot{H} \right)
F^{\prime\prime}(R)
+36\left(4H\dot H + \ddot H\right)^2 F^{\prime\prime\prime}(R) = 0\,.
\label{eq:3.12}
\end{eqnarray}

We now investigate the case in which $F(R)$ is given by 
$
F(R) = c_1 M^2 \left( R/M^2 \right)^{-n}
$, 
where $c_1$ is a dimensionless constant and $M$ denotes a mass scale. 
The form of $F(R)$ in Eq.~(\ref{eq:I8}) corresponds to the above expression 
with $n=-7/2$. 
In this case, the scale factor 
$a(t)$ and the scale curvature $R$ are given by~\cite{Briscese:2006xu}
$
a(t) = \bar{a} \left( t_s - t \right)^{(n+1)(2n+1)/(n+2)} 
$
and
$
R = 6n(n+1)(2n+1)(4n+5)(n+2)^{-2}\left( t_s - t \right)^{-2}
$, 
respectively. 
It follow from $d\hat{t} = \pm e^{\sigma/2} dt$ that the relation between 
the cosmic time in the Einstein frame $\hat{t}$ and one in the 
Jordan frame is given by 
$
\hat{t} = \mp \sqrt{-n c_1} \left(n+2\right)^n 
\left[ 6n(n+1)(2n+1)(4n+5) \right]^{-(n+1)/2} 
M^{n+1} \left( t_s - t \right)^{n+2}
$. 
If $n<-2$, the limit of $t \to t_s$ corresponds to that of
$\hat{t} \to \mp \infty$. For the case of Eq.~(\ref{eq:I8}),
$n =-7/2$. Thus, we see that the Big Rip singularity does not
appear in finite time for the scalar field theory,
although it emerges in the corresponding modified gravitational theory. 
The metric in the Einstein frame is expressed as 
$
d\hat{s}^2 = e^\sigma ds^2 = -d\hat{t}^2 + \hat{a}\left( \hat{t} \right)
d{\Vec{x}}^2
$, 
where $\hat{a}(t)$ is the scale factor in the scalar field theory given by 
$
\hat{a}(t) = \hat{\bar{a}}
\hat{t}^{3\left[\left(n+1\right)/\left(n+2\right) \right]^2} 
$, where $\hat{\bar{a}}$ is a constant. 
For $n =-7/2$, because when $t \to t_s$, $\hat{t} \to \mp \infty$, 
the scale factor in the scalar field theory $\hat{a}(t)$ diverges at infinite 
time. 
Consequently, the (finite-time) Big Rip singularity in $F(R)$ gravity is
transformed into the infinite-time singularity in the scalar field theory.
This shows the physical difference of late-time cosmological evolutions 
between two mathematically equivalent theories.

\section{Finite-time future singularities in $F(R)$ gravity \label{Sec3}}

In this section, we examine several models of $F(R)$-gravity with 
accelerating cosmological solutions ending at the finite-time future 
singularities by using the reconstruction technique explained in the preceding 
section. 

First, we consider the case of the Big Rip 
singularity~\cite{McInnes:2001zw}, in which $H$ behaves as 
$
H=h_0/\left(t_s - t\right)
$. 
Here, $h_0$ and $t_s$ are positive constants and $H$ diverges at $t=t_s$. 
In this case, 
if we neglect the contribution from the matter, 
the general solution of (\ref{eq:2.11}) is given by 
\be
\label{frlv11}
P(\phi) = P_+ \left(t_s - \phi\right)^{\alpha_+}
+ P_- \left(t_s - \phi\right)^{\alpha_-}\ ,\quad
\alpha_\pm \equiv \frac{- h_0 + 1 \pm \sqrt{h_0^2 - 10h_0 +1}}{2}\ ,
\ee
when $h_0 > 5 + 2\sqrt{6}$ or $h_0 < 5 - 2\sqrt{6}$ and 
\be
\label{rlv12}
P(\phi) = \left(t_s - \phi \right)^{-(h_0 + 1)/2}
\left( A_1 \cos \left( \left(t_s - \phi \right) \ln 
\frac{ - h_0^2 + 10 h_0 -1}{2}\right)
+ B_1 \sin \left( \left(t_s - \phi \right) \ln 
\frac{ - h_0^2 + 10 h_0 -1}{2}\right) \right)\ ,
\ee
when $5 + 2\sqrt{6}> h_0 > 5 - 2\sqrt{6}$. 
Here, $P_+$, $P_-$, $A_1$ and $B_1$ are constants. 
Using Eqs.~(\ref{eq:2.3}), (\ref{eq:2.4}) and (\ref{eq:2.12}), we find that 
when $R$ is large, the form of $F(R)$ is given by 
$
F(R) \propto R^{1 - \alpha_-/2}
$ 
for $h_0 > 5 + 2\sqrt{6}$ or $h_0 < 5 - 2\sqrt{6}$ case and 
$ 
F(R) \propto R^{\left(h_0 + 1\right)/4} \times \left(\mbox{oscillating parts}\right)
$ 
for $5 + 2\sqrt{6}> h_0 > 5 - 2\sqrt{6}$ case.

Next, we investigate more general singularity 
$
H \sim h_0 \left(t_s - t\right)^{-\beta}
$~\cite{Nojiri:2008fk}, 
where $h_0$ and $\beta$ are constants, and $h_0$ is assumed to be positive and 
$t<t_s$ because it should be for the expanding universe. 
Even for non-integer $\beta<0$, some derivative of $H$ and therefore
the curvature becomes singular. 
We should also note that in this case the scale factor $a$ 
behaves as 
$
a \sim \exp \left\{ \left[h_0/\left(\beta-1\right)\right] 
\left(t_s - t\right)^{-\left( \beta-1 \right)} 
+ \cdots \right\}
$, 
where $\cdots$ expresses the regular terms. From this expression, we find that 
if $\beta$ could not be any integer, the value of $a$, and therefore the value 
of the metric tensor, would 
become complex number and include the imaginary part when 
$t>t_s$, which is unphysical. This could tell that the universe could end at 
$t=t_s$ even if $\beta$ could be negative or less than $-1$.
We assume $\beta\neq 1$ because 
the case $\beta=1$ corresponds to 
the Big Rip singularity, which has been investigated. 
Furthermore, because the case $\beta=0$ corresponds to de Sitter space, 
which has no singularity, we take $\beta\neq 0$.
When $\beta>1$, the scalar curvature $R$ behaves as 
$
R \sim 12 H^2 \sim 12h_0^2 \left( t_s - t \right)^{-2\beta}
$. 
On the other hand, when $\beta<1$, the scalar curvature $R$ behaves as 
$
R \sim 6\dot H \sim 6h_0\beta \left( t_0 - t \right)^{-\beta-1}
$. 
We may obtain the asymptotic solution for $P$ when $\phi\to t_s$: 
(i) 
For $\beta>1$, 
$
P(\phi) \sim \exp \left\{ \left[ h_0/2\left(\beta - 1\right)\right] 
\left(t_s - \phi\right)^{-\beta + 1} \right\} 
\left(t_s - \phi\right)^{\beta/2}
\left(A_2 \cos \left(\omega \left(t_s - \phi\right)^{-\beta + 1}\right) 
+ B_2 \sin \left(\omega \left(t_s - \phi\right)^{-\beta + 1}\right) 
\right)
$, 
where 
$
\omega \equiv h_0/\left[2\left(\beta - 1\right)\right] 
$, and $A_2$ and $B_2$ are constants. 
When $\phi\to t_s$, $P(\phi)$ tends to vanish. 
(ii) 
For $1 > \beta > 0$, 
$
P(\phi) \sim B_3 \exp \left\{ -\left[ h_0/2\left(1 - \beta\right) \right] 
\left(t_s - \phi\right)^{1-\beta} \right\}
\left(t_s - \phi\right)^{\left(\beta + 1\right)/8}
$, 
where $B_3$ is a constant. 
(iii) 
For $\beta<0$, 
$
P(\phi) \sim A_3 \exp \left\{ -\left[ h_0/2\left(1 - \beta\right) \right] 
\left(t_s - \phi\right)^{1-\beta} \right\}
\left(t_s - \phi\right)^{- \left(\beta^2 - 6\beta + 1\right)/8}
$, 
where $A_3$ is a constant. 
Using Eqs.~(\ref{eq:2.3}), (\ref{eq:2.4}) and (\ref{eq:2.12}), 
we find the behavior of $F(R)$ (at large $R$) as summarized in Table~II. 

In the above investigations, we found the behavior of the scalar curvature $R$ 
from that of $H$. Conversely, we now consider the behavior of $H$ from that of 
$R$. When $R$ evolves as 
$
R \sim 6\dot H \sim R_0 \left( t_s - t \right)^{-\gamma}
$, 
if $\gamma>2$, which corresponds to $\beta = \gamma/2 >1$, 
$H$ behaves as 
$
H \sim \sqrt{ R_0/12 } \left( t_0 - t \right)^{-\gamma/2}
$, 
if $2>\gamma>1$, which corresponds to $1> \beta = \gamma -1 >0$, 
$H$ is given by 
$
H \sim \left\{ R_0/\left[6\left( \gamma - 1\right)\right] \right\} 
\left( t_s - t \right)^{-\gamma + 1}
$, 
and if $\gamma<1$, which corresponds to $\beta = \gamma -1 <0$, we 
obtain 
$
H \sim H_0 + \left\{ R_0/\left[ 6\left( \gamma - 1\right) \right] \right\} 
\left( t_s - t \right)^{-\gamma + 1}
$, where $H_0$ is an arbitrary constant and it does not affect 
the behavior of $R$. 
$H_0$ is chosen to vanish in 
$
H \sim h_0 \left(t_s - t\right)^{-\beta}
$. 
If $\gamma>2$, we find 
$
a(t) \propto \exp \left[ \left(2/\gamma -1 \right) 
\sqrt{R_0/12} \left( t_s - t \right)^{-\gamma/2 + 1} \right]
$, 
when $2>\gamma>1$, $a(t)$ behaves as 
$
a(t) \propto \exp \left( 
\left\{ R_0/\left[6\gamma\left( \gamma - 1\right)\right] \right\} 
\left( t_s - t \right)^{-\gamma}
\right)
$, 
and if $\gamma<1$, 
$
a(t) \propto \exp \left( H_0 t + 
\left\{ R_0/\left[ 6\gamma\left( \gamma - 1\right) \right] \right\} 
\left( t_s - t \right)^{-\gamma}\right) 
$. 
In any case, there appears a sudden future singularity~\cite{sudden} at
$t=t_s$. 

Since the second term in 
$
H \sim H_0 + \left\{ R_0/\left[ 6\left( \gamma - 1\right) \right] \right\} 
\left( t_s - t \right)^{-\gamma + 1}
$ 
is smaller than the first one, we 
may solve Eq.~(\ref{eq:2.11}) asymptotically as 
$ 
P\sim P_0 \left\{ 1 + \left[2h_0/\left(1-\beta\right)\right] 
\left(t_s - \phi\right)^{1-\beta} \right\} 
$ 
with a constant $P_0$, which gives 
$
F(R) \sim F_0 R + F_1 R^{2\beta/\left(\beta + 1\right)}
$, 
where $F_0$ and $F_1$ are constants. 
When $0>\beta>-1$, we find $2\beta/\left(\beta + 1\right)<0$.
On the other hand, when $\beta<-1$, we obtain
$2\beta/\left(\beta + 1\right)>2$. 
As we saw in 
$
F(R) \propto R^{1 - \alpha_-/2}
$ above, 
for $\beta<-1$, $H$ diverges when $t\to t_s$. 
Since we reconstruct $F(R)$ so that the behavior of $H$ could be recovered,
the $F(R)$ generates the Big Rip singularity when $R$ is large.
Thus, even if $R$ is small, the $F(R)$ generates a singularity
where higher derivatives of $H$ diverge.

We assume that $H$ behaves as 
$
H \sim h_0 \left(t_s - t\right)^{-\beta}
$. 
For $\beta> 1$, when $t\to t_s$,
$a\sim \exp \left[ h_0\left( t_s - t \right)^{1-\beta}/\left( \beta -1 \right) 
\right] \to \infty$ and
$\rho_{\rm eff} ,\, |p_{\rm eff}| \to \infty$.
If $\beta=1$, we find 
$a\sim \left(t_s - t\right)^{-h_0} \to \infty$ and
$\rho_{\rm eff} ,\, |p_{\rm eff}| \to \infty$.
If $0<\beta<1$, $a$ goes to a constant but $\rho ,\, |p| \to \infty$.
If $-1<\beta<0$, $a$ and $\rho$ vanish but $|p_{\rm eff}| \to \infty$.
When $\beta<0$, instead of 
$
H \sim h_0 \left(t_s - t\right)^{-\beta}
$, 
one may assume 
$
H \sim H_0 + h_0 \left(t_s - t\right)^{-\beta}
$. 
Hence, if $-1<\beta<0$, $\rho_{\rm eff}$ has a finite value $3H_0^2/\kappa^2$
in the limit $t\to t_s$.
If $\beta<-1$ but $\beta$ is not any integer, $a$ is finite and
$\rho_{\rm eff}$ and $p_{\rm eff}$ vanish if $H_0=0$ or $\rho_{\rm eff}$
and $p_{\rm eff}$ are finite if $H_0\neq 0$ but higher derivatives of $H$
diverge.
We should note that the leading behavior of the scalar curvature $R$ does not
depend on $H_0$ in 
$
H \sim H_0 + h_0 \left(t_s - t\right)^{-\beta}
$, 
and that the second term in this expression is 
relevant to the leading behavior of $R$. We should note, however, that $H_0$
is relevant to the leading behavior of the effective energy density
$\rho_{\rm eff}$ and the scale factor $a$.
\begin{table}[tbp]
\caption{Finite-time future singularities. 
Type I includes the case of $\rho$ and $p$ being finite at $t_s$. 
In case of Type IV, higher derivatives of $H$ diverges. 
Type IV also includes the case in which $p$ $(\rho)$ or both of them tend to 
some finite values while higher derivatives of $H$ diverge. 
Here, $t_s$ is the time when a singularity appears and 
$a_s$ is the value of $a(t)$ at $t=t_s$. 
}
\begin{center}
\begin{tabular}
{llccc}
\hline
\hline
Type
& Limit
& $a$ 
& $\rho$
& $|p|$
\\[0mm]
\hline
Type~I (``Big Rip")
& $t \rightarrow t_s$
& $a \rightarrow \infty$
& $\rho \rightarrow \infty$
& $|p| \rightarrow \infty$
\\[0mm]
Type~II (``sudden")
& $t \rightarrow t_s$
& $a \rightarrow a_s$
& $\rho \rightarrow \rho_s$
& $|p| \rightarrow \infty$
\\[0mm]
Type~III 
& $t \rightarrow t_s$
& $a \rightarrow a_s$
& $\rho \rightarrow \infty$
& $|p| \rightarrow \infty$
\\[0mm]
Type~IV 
& $t \rightarrow t_s$
& $a \rightarrow a_s$
& $\rho \rightarrow 0$
& $|p| \rightarrow 0$
\\[1mm]
\hline
\hline
\end{tabular}
\end{center}
\label{tb:table1}
\end{table}
\begin{table}[tbp]
\caption{Summary of the behavior of $F(R)$ gravity in case of 
$H \sim h_0 \left(t_s - t\right)^{-\beta}$. Here, 
$c_1= \left[ h_0/2\left(\beta - 1\right) \right] 
\left(12h_0\right)^{-(\beta - 1)/\left(2\beta\right)}$ 
and 
$c_2= \left[ h_0/2\left(1-\beta\right) \right] 
\left(-6h_0\beta\right)^{(\beta - 1)/\left(\beta+1\right)}$. 
We note that $-6h_0 \beta R >0$ when $h_0, R>0$. 
}
\begin{center}
\begin{tabular}
{ccccc}
\hline
\hline
& Type~I (``Big Rip")
& Type~II (``sudden")
& Type~III 
& Type~IV 
\\[0mm]
\hline
$\beta$
& $\beta > 1$
& $-1<\beta<0$
& $0<\beta<1$
& $\beta<-1,\,\beta:\mbox{not integer}$
\\[0mm]
$F(R)$ 
& 
$F(R)\propto e^{c_1 R^{\frac{\beta -1}{2\beta}}} R^{-\frac{1}{4}}$
& 
$F(R)\propto e^{-c_2 R^{\frac{\beta -1}{\beta+1}}}
R^{\frac{\beta^2 + 2\beta + 9}{8(\beta +1)}}$
&
$F(R)\propto e^{-c_2 R^{\frac{\beta -1}{\beta+1}}} R^{\frac{7}{8}}$
&
$F(R)\propto e^{-c_2 R^{\frac{\beta -1}{\beta+1}}}
R^{\frac{\beta^2 + 2\beta + 9}{8(\beta +1)}}$
\\[1mm]
\hline
\hline
\end{tabular}
\end{center}
\label{tb:table2}
\end{table}

In Ref.~\cite{Nojiri:2005sx}, the finite-time future 
singularities has been classified as shown in Table~I.  
The Type I corresponds to $\beta>1$ or $\beta=1$ case, Type II to $-1<\beta<0$ 
case, Type III to $0<\beta<1$ case, and Type IV to $\beta<-1$ but $\beta$ is 
not any integer number. 
Thus, we have constructed several examples of $F(R)$ gravity showing 
the above finite-time future singularities of any type. 
It also follows from the reconstruction method that 
there appears Type I singularity 
for $F(R) = R + \tilde{\alpha} R^n$ with $n>2$ and Type III singularity for
$F(R) = R - \tilde{\beta} R^{-n}$ with $n>0$, where $\tilde{\alpha}$ and
$\tilde{\beta}$ are constants. 
In fact, however, even if some specific model contains the finite-time 
future singularity, one can always reconstruct the model 
in the remote past in such a way that the finite-time future singularity could 
disappear. Positive powers of the curvature (polynomial structure) usually 
help to make the effective quintessence/phantom phase become transient and 
to avoid the finite-time future singularities. 
The corresponding examples have been examined in 
Refs.~\cite{Nojiri:2008fk, Abdalla:2004sw}.

\section{Conclusion}

In the present article, we have reviewed finite-time future singularities 
in modified gravity. 
We have reconstructed an explicit model of modified gravity realizing a 
crossing of the phantom divide. 
It has been shown that 
the Big Rip singularity appears in this modified gravitational theory, 
whereas that the (finite-time) Big Rip singularity in the modified gravity is 
transformed into the infinite-time singularity in the corresponding scalar 
field theory. In addition, we have examined several models of 
modified gravity which predict accelerating cosmologies ending at the 
finite-time future singularities of all four known types.

\section*{Acknowledgments} 
The author deeply appreciates the invitation of Professor Sergei D. Odintsov 
to submit this article to the special volume 
\textit{Casimir effect and Cosmology} on the occasion of the 70th birthday of 
Professor I. Brevik 
published by Tomsk State Pedagogical University, 2008. 
He also thanks Professor Chao-Qiang Geng, Professor Shin'ichi Nojiri and 
Professor S.~D.~Odintsov for their collaboration in 
Refs.~\cite{Bamba:2008hq, Bamba:2008ut} very much. 
In addition, he is grateful to Professor Misao Sasaki for very helpful 
discussion of related problems. 
This work is supported in part by 
National Tsing Hua University under Grant \#: 97N2309F1.


\end{document}